\documentstyle[12pt]{article}
\title{\bf Classical and quantum wormholes in a flat $\Lambda$-decaying cosmology }
\author{F. Darabi \thanks{This work has been financially supported by Research Institute for
Fundamental Sciences, Tabriz, Iran.} \\
{\small Department of Physics, Azarbaijan University of Tarbiat
Moallem, 53714-161 Tabriz, Iran .}\\
{\small Research Institute for Fundamental Sciences, Tabriz, Iran.
 }}
\begin{document}
\maketitle
\begin{abstract}
We study the classical and quantum wormholes for a flat {\it
Euclidean} Friedmann-Robertson-Walker metric with a perfect fluid
including an ordinary matter source plus a source playing the role
of dark energy (decaying cosmological term). It is shown that
classical wormholes exist for this model and the quantum version
of such wormholes are consistent with the Hawking-Page conjecture
for quantum wormholes as solutions of the Wheeler-DeWitt equation.
\end{abstract}

\newpage

\section{Introduction}

Classical wormholes are usually considered as Euclidean metrics
that consist of two asymptotically flat regions connected by a
narrow throat (handle). Wormholes have been studied mainly as
instantons, namely solutions of the classical Euclidean field
equations. In general, such wormholes can represent quantum
tunneling between different topologies. They are possibly useful
in understanding black hole evaporation \cite{Haw}; in allowing
nonlocal connections that could determine fundamental constants;
and in vanishing the cosmological constant $\Lambda$ \cite{Col}.
They are even considered as an alternative to the Higgs mechanism
\cite{Mig}. Consequently, such solutions are worth finding.

Unfortunately, they exist for certain special kinds of matter
\cite{Gid} . For example, they exist for an imaginary minimally
coupled scalar field \cite{HP}, but do not exist for pure gravity.
Due to limited known classical wormhole solutions, Hawking and
page advocated a different approach in which wormholes were
regarded, not as solutions of the classical Euclidean field
equations, but as solutions of the quantum mechanical
Wheeler-DeWitt equation \cite{HP}. These wave functions have to
obey certain boundary conditions in order that they represent
wormholes. The boundary conditions seem to be that: (i) the wave
function is exponentially damped for large tree geometries; (ii)
the wave function is regular in some suitable way when the
tree-geometry collapses to zero.

The first condition express the fact that spacetime should be
Euclidean when tree geometries become infinite. The second
condition should reflect the fact that spacetime is nonsingular
when tree geometries degenerates, namely the wave function should
not oscillate an infinite number of time.

Therefore, in general, an open and interesting problem is whether
classical and quantum wormholes can occur for fairly general
matter sources. To the author's knowledge, the dark energy as a
$\Lambda$-decaying source has not been received much attention, in
this regard. Although there are few works attempting to study such
a universe (in a different formulation), using quantum effective
action \cite{Odi}, we shall consider this model and seek for the
classical and quantum wormhole solutions, explicitly.

\section{Classical wormholes}

The reason why classical wormholes may exist is related to the
implication of a theorem of Cheeger and Glommol \cite{Chee} which
states that a {\it necessary} condition for a wormhole to exist is
that the eigenvalues of the Ricci tensor be negative {\it
somewhere} on the manifold. This is necessary but not {\it
sufficient} condition for their existence. For example, the
energy-momentum tensors of an axion field and of a conformal
scalar field are such that, when coupled to gravity, the Ricci
tensor has negative eigenvalues. However, for a minimally coupled
scalar field with the lagrangian density
$$
{\cal L}=\frac{1}{2}\nabla_c\phi \nabla^c\phi,
$$
and the energy-momentum tensor
$$
T_{ab}=\nabla_a\phi
\nabla_b\phi-\frac{1}{2}g_{ab}\nabla_c\phi\nabla^c\phi,
$$
the Einstein equations read
$$
R_{ab}=\nabla_a\phi \nabla_b\phi,
$$
which shows $R_{ab}$ can never be negative.

Taking this into account, we consider an Euclidean
Friedmann-Robertson-Walker (FRW) metric
\begin{equation}
ds^2=dt^2+a^2(t)\left[\frac{dr^2}{1-kr^2}+r^2(d\theta^2+\sin^2\theta
d\phi^2)\right], \label{1}
\end{equation}
where $a(t)$ is the scale factor and $k=0, +1$ and $-1$ account
for flat, closed and open universes, respectively. This metric
evolves according to the Einstein equation \footnote{We have used
the units where $8\pi G=c=1$.}
\begin{equation}
{\cal R}_{\mu \nu}-\frac{1}{2}g_{\mu \nu}{\cal R}=T_{\mu \nu},
\label{2}
\end{equation}
where we take the energy momentum tensor $T_{\mu \nu}$ to be
perfect fluid
\begin{equation}
T_{\mu \nu}=(\rho+p)u_{\mu} u_{\nu}+pg_{\mu \nu}. \label{3}
\end{equation}
The time-time and space-space components of the Einstein equation
(\ref{2}) leads respectively to
\begin{equation}
\frac{\dot{a}^2}{a^2}=\frac{k}{a^2}-\frac{\rho}{3}, \label{4}
\end{equation}
\begin{equation}
2\frac{\ddot{a}}{a}+\frac{\dot{a}^2}{a^2}-\frac{k}{a^2}=p,
\label{5}
\end{equation}
where Eq.(\ref{4}) is the Euclidean Friedmann equation in which
the derivative is with respect to the Euclidean time. There is
also a conservation equation $\nabla_{\mu}T^{\mu \nu}=0$ whose
time component gives the fluid equation
\begin{equation}
\dot{\rho}+3\frac{\dot{a}}{a}(\rho+p)=0. \label{7}
\end{equation}
Wormholes are typically described by the Euclidean Friedmann
equation (\ref{4}) with a scale factor dependent density,
typically as $\rho \sim a^{-n}$. We now assume a {\it flat} FRW
universe with a perfect fluid source combined of ordinary matter
and a source evolving with the scale factor
\begin{equation}
\rho=\rho_m+\rho_v=\left(\frac{\rho_0}{a^{3\gamma}}-\frac{\Lambda_0}{a^2}\right).
\label{10}
\end{equation}
The first term is the ordinary matter density and the second term is
a density playing the role of (negative) dark energy - a decaying
cosmological term. In fact, there are theoretical and observational
motivations for considering models in which $\Lambda$ decays. For
example, such behaviors as $\Lambda \sim t^{-2}$ \cite{Orf} or
$\Lambda \sim a^{-m}$ \cite{SW} have already been reported. For
$0\leq m<3$ \cite{SW}, the effect of decaying cosmological term on
the cosmic microwave background anisotropy is studied and the
angular power spectrum for different values of $m$ and density
parameter $\Omega_{m 0}$ is computed. Models with $\Omega_{m 0}\geq
0.2$ and $m \geq 1.6$ are shown to be in good agreement with data.
For $m=2$ \cite{CW}, it is shown that in the early universe
$\Lambda$ could be several tens of orders bigger than its present
value, but not big enough disturbing the physics in the
radiation-dominant epoch in the standard cosmology. In the
matter-dominant epoch such a varying $\Lambda$ shifts the three
space curvature parameter $k$ by a constant which changes the
standard cosmology predictions reconciling observations with the
inflationary scenario. Such a vanishing cosmological term also leads
to present creation of matter with a rate comparable to that in the
steady-state cosmology \cite{CW}.

To obtain the necessary condition for the existence of wormholes
we evaluate the eigenvalues of the Ricci tensor
\begin{equation}
R_{\mu \nu}=T_{\mu \nu}-\frac{1}{2}g_{\mu \nu}T,
\end{equation}
which together with the perfect fluid (\ref{3}) and the equation
of state (\ref{13}) ( see below ) leads, for example, to
\begin{equation}
R_{00}=\frac{\rho_0}{a^{3\gamma}}\left[1-\frac{\gamma}{2}\right]-\frac{4}{3}\frac{\Lambda_0}{a^2}.
\label{12}
\end{equation}
It is seen that $R_{00}$ can be negative at large $a$ for any
$\gamma>2/3$. However, it does not guarantee that wormholes
certainly do exist for this range of $\gamma$. It is therefore,
necessary to check out the {\it sufficient} condition.

Substitution for $\rho$ from (\ref{10}) and $k=0$ into
Eq.(\ref{4}) leads to\footnote{Recent observations strongly
indicate that the universe is almost flat, namely $k=0$.}
\begin{equation}
\frac{\dot{a}^2}{a^2}=\frac{1}{3}\left(\frac{\Lambda_0}{a^2}-\frac{\rho_0}{a^{3\gamma}}\right).
\label{11}
\end{equation}
In order to have an asymptotically Euclidean wormhole $\dot{a}^2$
must remains positive at large $a$. This wormhole then represents
two separate asymptotically Euclidean regions joined together by a
throat with the finite size $a_0$ at which $\dot{a}=0$. Since
according to {\it strong energy condition} $\rho_0>0$, then we can
not have this wormhole for $\Lambda_0<0$ because $\dot{a}^2$
becomes negative. Therefore, classical Euclidean wormholes are
possible for $\Lambda_0>0$, namely if the dark energy has a
negative energy density $\rho_v=-\frac{\Lambda_0}{a^2}<0$. On the
other hand, even for $\Lambda_0>0$ we need $\gamma>\frac{2}{3}$ to
have a positive $\dot{a}^2$ at large $a$.

To determine the equation of state, we substitute for $\rho$ from
Eq.(\ref{10}) in the conservation equation (\ref{7}) and obtain
the following equation of state
$$
p=p_m+p_v
$$
\begin{equation}
\hspace{20mm}=\rho_m(\gamma-1)-\frac{1}{3}\rho_v, \label{13}
\end{equation}
where the first term describes the standard equation of state for
ordinary matter and the second term accounts for equation of state
for the dark energy.

In conclusion, considering the {\it necessary} and {\it
sufficient} conditions as $\gamma>\frac{2}{3}$, one finds that
classical Euclidean wormholes are possible for the matter source
(\ref{10}) if and only if $\gamma> 2/3$, provided the dark energy
has a negative energy density. In the Lorentzian sector of the
model, it is easy to show that $\gamma>\frac{2}{3} $ leads the
strong energy condition to hold for the total pressure $p$ and
total density $\rho$.

\section{Quantum Wormholes}

As a matter of fact, the number of known classical wormholes is so
limited. It casts doubt on whether wormholes are important, only
in the very restricted class of theories, in which the matter
content allows wormhole instantons? To resolve this problem,
Hawking and Page advocated a different approach and considered
that solutions of the Wheeler-DeWitt equation could more generally
represent the wormholes \cite{HP}. They realized that for the mini
superspace models one may consider metrics of the Euclidean
Friedmann form
\begin{equation}
ds^2=N^2(t)dt^2+a^2(t)d\Omega^2_3.
\end{equation}
If $N$ is imaginary, this is the Lorentzian metric, and if $N$ is
real, it is the metric of an Euclidean wormhole. However,
solutions of the Wheeler-DeWitt equation are independent of the
lapse function $N$ and $t$. So, they can be interpreted either as
Friedmann universe, or as wormholes according to the appropriate
boundary conditions. The boundary conditions for wormholes seems
to be that the wave functions should decay exponentially for large
scale factor $a$, so as to represent Euclidean space, and that
they be regular in some suitable way as $a\rightarrow 0$, so that
no singularities are present.

By defining $R=\sqrt{\frac{3}{\Lambda_0}}a$ we obtain the rescaled
Friedmann equation
\begin{equation}
\frac{\dot{R}^2}{R^2}=\frac{1}{R^2}-\frac{\alpha_0}{R^{3\gamma}},
\label{12}
\end{equation}
with the constant
\begin{equation}
\alpha_0=\frac{\rho_0}{(\frac{3}{\Lambda_0})^{3\gamma/2}}.
\label{12'}
\end{equation}
The quantum mechanical version of this equation is given by
\cite{Hal}
\begin{equation}
\left(R^2\frac{d^2}{dR^2}+qR\frac{d}{dR}+\alpha_0 R^{6-3\gamma}-R^4
\right)\Psi(R)=0, \label{14}
\end{equation}
where $q$ represents part of the factor ordering ambiguities. We
set $q=0$ and study the potential to get some idea as to when a
Euclidean domain occurs at large $R$ by considering the sign of
the potential
\begin{equation}
U(R)=\alpha_0 R^{4-3\gamma}-R^2,
\end{equation}
in the equation
\begin{equation}
\left[\frac{d^2}{dR^2}+U(R)\right]\Psi(R)=0. \label{15}
\end{equation}
For positive potential $U(R)>0$, oscillating solutions occur which
represent Lotentzian metrics. On the contrary, for negative
potential $U(R)<0$, wormhole solutions can occur which are
asymptotically Euclidean at large $R$. The potential is negative
for $\gamma>2/3$ in the case of positive energy density $\rho_0$.
Therefore, wormholes obeying the Hawking-Page boundary condition
at large $R$, occur when $\gamma>2/3$ is valid for the source
(\ref{10}). The presence of any matter source with $\gamma<2/3$
will eventually dominate for large $R$ and prevent the
asymptotically Euclidean wormholes to occur.

Asymptotically Euclidean property of the wave function is not
sufficient to make it a wormhole. It also requires regularity for
small $R$. In order to realize this, we can ignore $R^4$ term in
Eq.(\ref{14}) as $R\rightarrow 0$, when $\gamma>2/3$. In this case,
the Wheeler-DeWitt equation (\ref{14}) (for $\gamma \neq 2$)
simplifies to a Bessel differential equation with solution
\begin{equation}
\Psi(R)\simeq R^{(1-q)/2}\left[c_1
J_{\nu}\left(\frac{2\sqrt{\alpha_0}}{3(2-\gamma)}R^{3-3\gamma/2}\right)+
c_2Y_{\nu}\left(\frac{2\sqrt{\alpha_0}}{3(2-\gamma)}R^{3-3\gamma/2}\right)\right],
\label{16}
\end{equation}
where use has been made of $\nu\equiv(1-q)/3(2-\gamma)$. The
wormhole boundary condition at $R\rightarrow 0$ is satisfied for the
Bessel function of the $J$ kind.

In the particular case $\gamma=2$, the solution of Eq.(\ref{14})
with $q=1$ is a linear combination of Bessel functions $J_{\pm
i\sqrt{\alpha_0}/2 (ia^2/2)}$ which oscillates an infinite number of
times at $R\rightarrow 0$ and therefore can not satisfy the required
regularity condition for a quantum wormhole.

On the other hand, in the case of $\gamma=4/3$ which represents
radiation ( or equivalently, that of a conformally coupled scalar
field ) dominated FRW ansatz, Eq.(\ref{14}) for $q=0$ is written
as
\begin{equation}
\left(\frac{d^2}{dR^2}+\alpha_0-R^2 \right)\Psi(R)=0, \label{17}
\end{equation}
which is in the form of a parabolic equation with solution in terms
of confluent hypergeometric functions \cite{Abr}
\begin{equation}
\Psi(R)\simeq \exp(-R^2/2)[c_3 \:_1F_1(\frac{1}{4}(1-\alpha_0); 1/2;
R^2)+c_4 \:_1F_1(\frac{1}{4}(3-\alpha_0); 3/2; R^2)].\ \label{18}
\end{equation}
For example, for $c_3=0$ with $\alpha_0=(35, 55)$, and $c_4=0$ with
$\alpha_0=(25, 37)$, we obtain regular oscillations at $R\rightarrow
0$, and Euclidean regimes for large $R$, see Figs.1, 2, 3, 4.

Therefore, Hawking-Page boundary conditions are satisfied for some
special values of $\alpha_0$ and so we have a spectrum of
wormholes. Considering Eq.(\ref{12'}), it turns out that for a
given equation of state with $\gamma>2/3$, the existence of
quantum wormholes depends on the special values of $\rho_o$ and
$\Lambda_0$. In other words, the spectrum of wormholes depends on
the spectrum of $\rho_o$ and $\Lambda_0$. We notice that there is
no such constraint on these values for the occurrence of classical
wormholes.

\section*{Conclusion}

The classical and quantum wormhole solutions have been studied for
a flat Euclidean Friedmann-Robertson-Walker metric coupled to a
perfect fluid combined of an ordinary matter source and a source
playing the role of dark energy ( a decaying cosmological term ).
We have shown that classical and quantum wormholes exist for this
matter source for which the strong energy condition could be hold
in the Lorentzian sector of the model. An spectrum of quantum
wormholes has been obtained, for each of which there is a specific
relation (\ref{12'}) between the characteristics of matter and
dark energy densities.

\section*{Acknowledgment}

This work has been financially supported by Research Institute for
Fundamental Sciences, Tabriz, Iran.
\\
\\
{\large{\bf Figure captions}}
\vspace{10mm}\\
Figure 1. A quantum wormhole solution for the case $\gamma=4/3$ with
$c_3=0$ and $\alpha_0=35$.\\
\\
Figure 2. A quantum wormhole solution for the case $\gamma=4/3$ with
$c_3=0$ and $\alpha_0=55$.\\
\\
Figure 3. A quantum wormhole solution for the case $\gamma=4/3$ with
$c_4=0$ and $\alpha_0=25$.\\
\\
Figure 4. A quantum wormhole solution for the case $\gamma=4/3$ with
$c_4=0$ and $\alpha_0=37$.


\newpage

\begin{thebibliography}{99}
\bibitem{Haw}S. W. Hawking, Phys. Rev. D.{\bf 37} (1988), 904.
\bibitem{Col}S. Coleman, Nucl. Phys. B{\bf 307} (1988), 864; B{\bf 310} (1988), 643;
I. Klebanov, L. Susskind, and T. Banks, Nucl. Phys. B{\bf 317} (1989), 665..
\bibitem{Mig}S. Mignemi and I. Moss, Phys. Rev. D.{\bf 48} (1993), 3725.
\bibitem{Gid}S. Giddings, A. Strominger, Nucl. Phys. B{\bf 306}
(1988), 890; K. Lee, Phys. Rev. Lett.{\bf 61} (1988), 263; A.
Hosoya, W. Ogura, Phys. Lett. B{\bf 225} (1989), 117; A. K. Gupta,
J. Hughes, J. Preskill, and M. B. Wise, Nucl. Phys. B{\bf 333}
(1990), 195; D. H.Coule and K. Maeda, Class. Quantum. Grav. {\bf
7} (1990), 955.
\bibitem{HP}S. Hawking, D. N. Page, Phys. Rev. D.{\bf 42} (1990), 2655.
\bibitem{Odi}S. Nojiri, O. Obregon, S. D. Odintsov and K. E. Osetrin, Phys. Lett. B{\bf 449} (1999),
173; Phys. Lett. B{\bf 458} (1999), 19: A. Carlini and M. Mijic,
SISSA preprint, 91A (1990); A. Carlini, D. H. Coule, and D. M.
Solomons, Mod. Phys. Lett. A.{\bf 11} (1996), 1453.
\bibitem{Chee}J. Cheeger and D. Grommol, Ann. Math. {\bf 96} (1972), 413.
\bibitem{Orf}O. Bertolami, Nuovo Cim. B.{\bf 93} (1986), 36.
\bibitem{SW} V. Silveira and I. Waga, Phys. Rev. D {\bf 56}, (1997), 4625;
Phys. Rev. D. {\bf 50} (1994), 4890.
\bibitem{CW} W. Chen and Y-S. Wu, Phys. Rev. D {\bf 41} (1990), 695 .
\bibitem{Zhu}A. Zhuk, Phys. Lett. A{\bf 176} (1993), 176.
\bibitem{Hal}J. J. Halliwell, in {\it Quantum Cosmology and Baby
Universes} (World Scientific, 1991).
\bibitem{Abr}M. Abramowitz and I. A. Stegun, {\it Handbook of
Mathematical functions} (Dover, 1965).
\end{thebibliography}
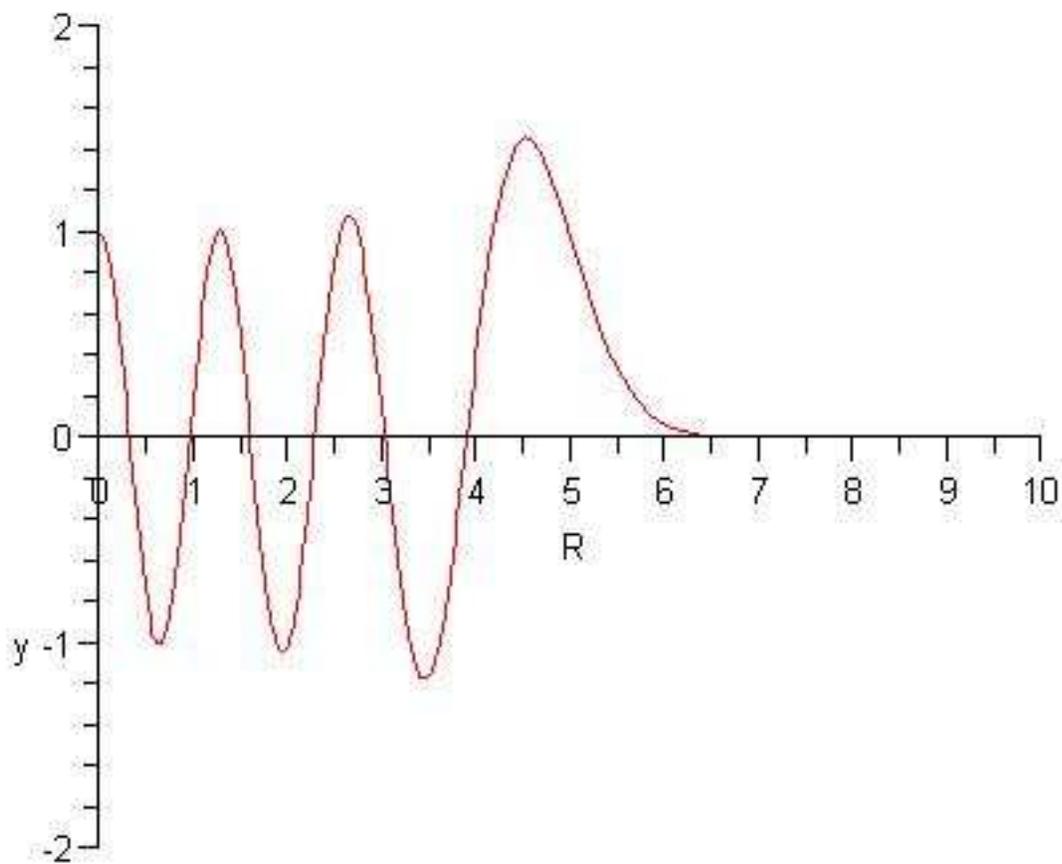
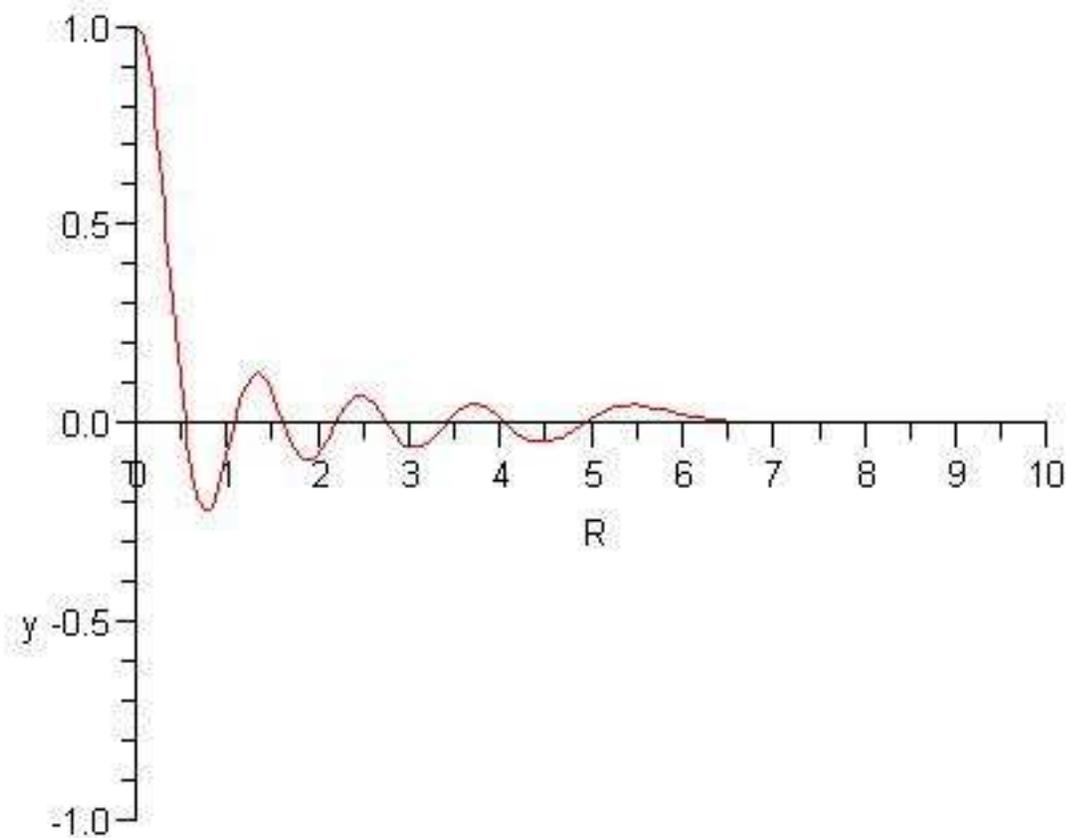
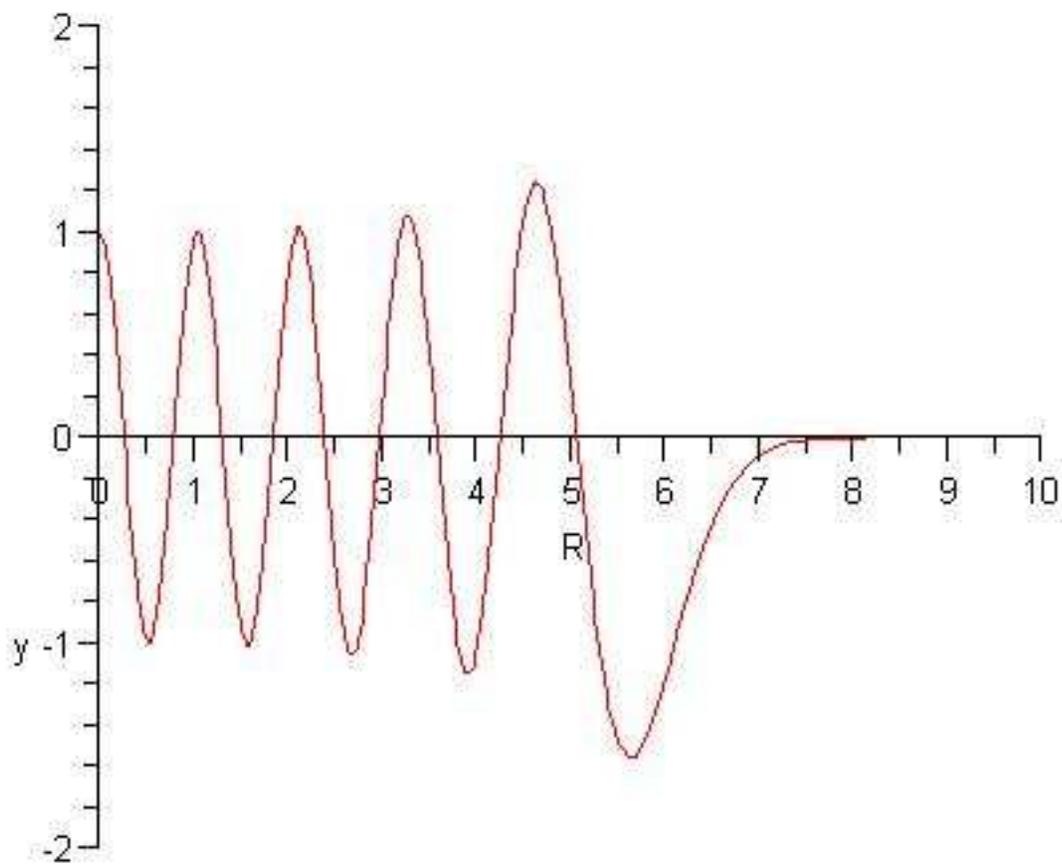
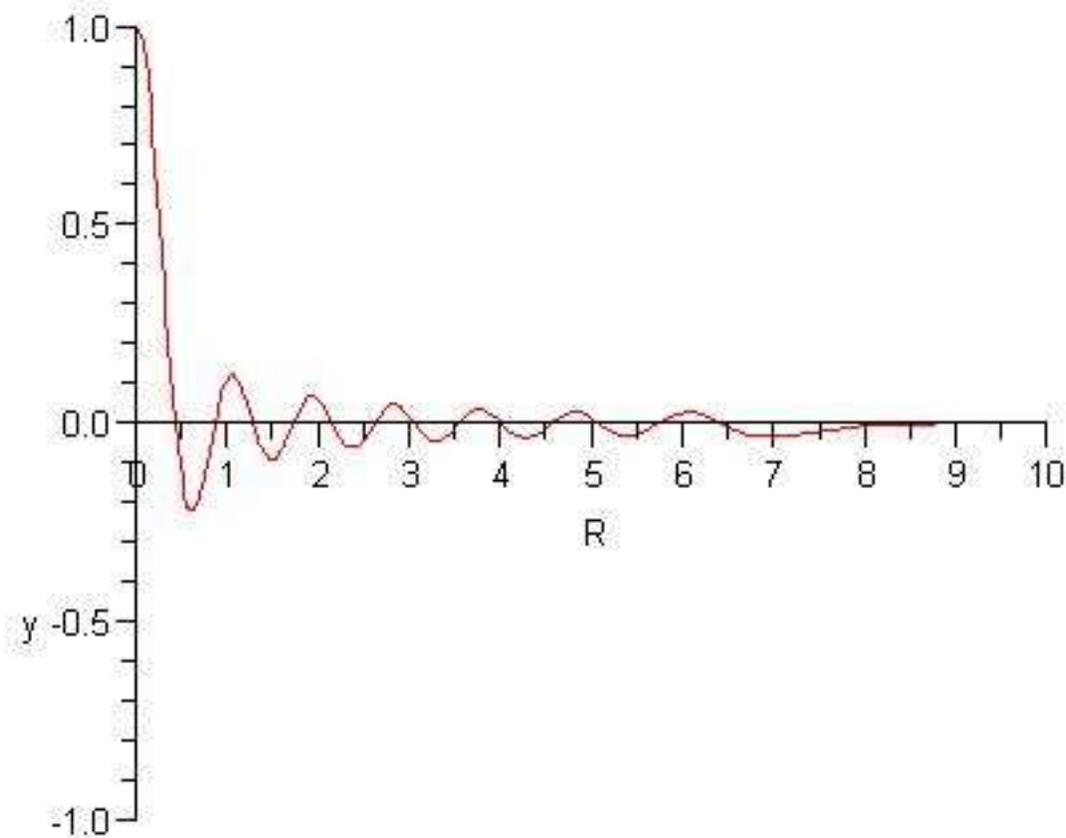
\end{document}